# Control of Noise in Chemical and Biochemical Information Processing

**Vladimir Privman**[a]

[a] *Department of Physics, Clarkson University*
*Potsdam, NY 13699, USA*
*tel: 1 315 268 3891*
*e-mail: privman@clarkson.edu*

---

**Abstract:** We review models and approaches for error-control in order to prevent the buildup of noise when gates for digital chemical and biomolecular computing based on (bio)chemical reaction processes are utilized to realize stable, scalable networks for information processing. Solvable rate-equation models illustrate several recently developed methodologies for gate-function optimization. We also survey future challenges and possible new research avenues.

**Keywords**: biocomputing · biomedical · biomolecular · computing · digital · sensor

---

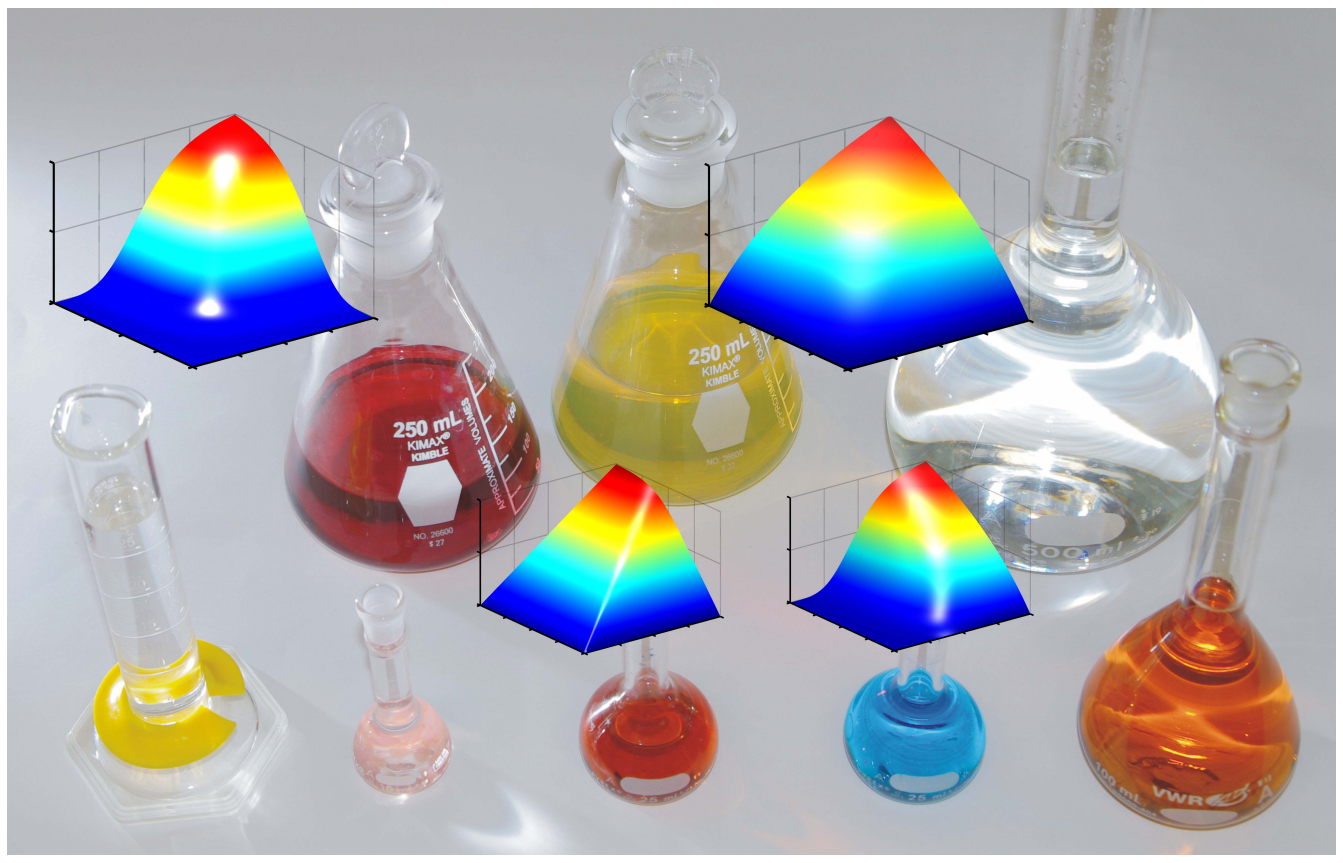

The desirable (top-left schematic) noise-suppressing but presently experimentally elusive, and the typical (top-right schematic) experimentally observed noise-amplifying **AND**-gate response surfaces for *(bio)chemical computing*. Two recently experimentally realized response surfaces which can be in the parameter regime of no noise amplification are also shown (lower schematics).

---

## 1. Introduction

Accompanying articles in this Special Issue review recent advances in the realizations of chemical[1-4] and biomolecular[5-12] systems which process information by binary gate functions, for example, **AND**, **OR**, etc., by utilizing (bio)chemical processes. The information processing and storing "units" need not be limited to molecules. They can include[1-13] natural and synthetic supra-molecular, biomolecular and/or catalytic structures (enzymes, DNA, etc.), as well as whole cells. Present chemical or biochemical "information processing," to be termed "computing" for brevity, systems have not been versatile and practical enough to compete with the conventional silicon (Si) computers. The short term applications for biocomputing have been envisaged in offering additional functionalities for multi-signal sensing[14,15] and interfacing/actuation[15-17] when excessive wiring to Si-computers and power sources is not desirable, e.g., in biomedical implants, diagnostic patches, etc.

Beyond the design of various binary gates, chemical and biochemical computing face the challenge of developing functionalities to connect the gates and other network components to enable fault-tolerant information processing of increasing complexity.[18-20] Reported results for networks[1-4,18-20] of (bio)chemical information processing units have included systems performing elements of basic arithmetic operations,[21-22] multifunctional molecules,[23-25] DNA-based gates and circuits,[26,27] and enzyme-catalyzed reactions realizing concatenated gates.[19,20,28,29]

Here we introduce, by illustrative model examples, concepts in noise reduction and control for scalability in biochemical computing. The approach has been tested in experimental realizations for enzyme-reaction based logic gates and networks.[10,12,14,17,19,30] However, the reported theoretical ideas generally apply to a broad range of chemical and biomolecular information processing systems, presently suggesting that typical networks of up to 10 binary gates can operate with the acceptable level of noise,[10,12,17] similar to findings in networking of neurons.[31,32] For networks of more than order 10 binary steps, additional non-binary network elements, as well as proper network design to utilize redundancy for digital error correction, will be needed for fault-tolerant operation.[10,12,17,30]

The level of noise in the environments envisaged for applications of future chemical[1-4] and biomolecular[5-12,14,17,19-20,31-34] computing systems is quite high as compared to the electronic computer counterparts. Indeed, both the input/output signals and the "gate machinery" chemical concentrations, can typically fluctuate several percent or more, on the scale normalized to the digital **0** to **1** range. Avoiding noise amplification by careful design and parameter selection for gates and networks, is therefore quite important even for relatively small networks. Here we do not address the *origin/sources* of stochastic and environmental noise in (bio)chemical reactions. We also do not consider the experimental findings, which are reviewed in other articles in this Special Issue. Instead, we survey solvable rate equation models which serve to illustrate recently developed concepts in (bio)chemical computing gate design for noise suppression.

Theoretical considerations reviewed here apply to numerous reported chemical and biochemical information processing systems. Indeed, chemical processes can be cast[35,36] in the language of computing operations, with signals represented by changes[1-4,36-48] in structural, chemical, or physical properties, resulting due to physical,[49-76] chemical,[77-84] or more than one type[53,85-87] of input. The output signals can be detected spectroscopically[88-93] or electrically/electrochemically.[94-96] Chemical computing can be done in the bulk, e.g., in solution,[97-98] or at surfaces/interfaces,[14-17,99-102[ such as at electrodes or on Si-chips. Supra-molecular structures have also been considered as switchable devices for logic operations.[103]

Chemical-computing equivalents of standard binary gates, such as **AND**,[104,105] **OR**,[106] **XOR**,[103,107] **NOR**,[91,93,108-111] **NAND**,[111-113] **INHIB**,[114-117] **XNOR**[118,119] have been realized. Reversibility,[120,121] reconfigurability[122-125] and resettability[126,127] of the resulting gates have been explored. Digital logic functions of several-gate device components[128-134] have been realized, such as keypad lock and memory units.[85,135-145] Chemical-computing systems can function at the single-molecule[146] nano-scale devices,[147] as well as perform parallel computations by numerous molecules.[148]

Chemical computing shows great promise,[149-151] though, as most unconventional computing approaches,[152] it is not being developed as an alternative to the speed and versatility of Si-computers. The short-term focus has been on novel functionalities and applications: microrobotics, multi-input (bio)sensors/actuators, implantable device components. The main challenge for chemical computing has been networking of basic gates for achieving scalable, fault-tolerant information processing networks.[153] Small networks performing basic operations have been realized,[1-4,21,22] for example, adder/subtractor and their sub-units.[154-159] Multi-signal response to chemical or physical inputs has been demonstrated,[23-25] and attempts at scaling up the complexity according to biological principles have been reported.[74]

In principle, biomolecular or *biochemical* information processing, to be termed "biocomputing" for brevity, constitutes a branch of chemical computing. However, it has drawn a lot of recent interest, because biomolecules offer natural specificity when used in complex "chemical soup" environments, as well as biocompatibility, the latter important for biomedical/biotechnology applications. Biomolecules are also more suitable for scalability paradigms borrowing ideas from nature. Furthermore, the biocatalytic nature of many utilized biomolecular processes, offers certain advantages for analog noise control in the binary-gate circuit design paradigm.[17] Proteins/enzymes[5-12,14-17,19-20,161-165] and DNA/RNA/DNAzymes[6,7,166-182] have been extensively used for gates, for realizing small networks and computational units, and for systems motivated[183] by applications.

This review is organized as follows. In Section 2, we describe general concepts for considering (bio)chemical binary gates. Gate design for decreasing noise amplification is addressed in Section 3. Section 4 addresses optimization of AND gates. Section 5 describes gate design as part of a network. Section 6 is devoted to summary and discussion of future challenges.

## 2. From (Bio)chemical Information Processing Gates to Networks

Processing of large quantities of information at high levels of complexity requires utilization of a paradigm of scalable networking of simple gates. Recent chemical and biochemical computing literature has usually implicitly assumed an approach similar to that used in Si-chip electronic devices[184,185] of designing fault-tolerant systems that can avoid buildup of noise without prohibitive use of resources. However, with biomolecules, one could perhaps also use design concepts borrowed from processes in living organisms.[186] Hybrid solutions can be expected, with bio-inspired elements supplementing the electronics designs. Other approaches include massive parallelism,[182] specifically with DNA.[187]

Thus far, the vast majority of the recent enzyme-based biocomputing realizations and analyses reported,[10,12,14-17,19,20,30] and similarly the rest of the biomolecular computing literature, have at best realized only small networks of gates, even though the aim has been to follow the digital approach based on analog gates and other elements connected in increasing-complexity networks.[184,185] Biomolecular computing is presently also far from the complexity of coupled biochemical reaction sets needed for mimicking processes in living organisms. Additionally, near-future applications of moderate-complexity biocomputing systems will likely be in novel sensor systems,[15] processing several input signals and yielding Yes/No digital outputs, corresponding to the "Sense/Respond" or "Sense/Diagnose/Treat" actions. Therefore, both the biochemical steps and the output transduction to electrodes/electronic computers for the "action" step, suggest the use of the binary Yes/No digitalization.

More importantly, the binary/digital information processing paradigm offers an approach for control of the level of noise buildup in complex networks. Chemical and biochemical systems operate as larger levels of noise than electronic computers. The inputs reactant and the "gate machinery" chemical concentrations, e.g., catalysts, typically fluctuate within at least a couple of percent of the range of values identified as the binary **0** and **1**, and careful attention to the control of noise build-up is required for networks as small as 2-3 gates.[10,15,19,20]

While we talk of digital information processing, the network elements are actually always analog. Figure 1 offers an illustration for the simplest "gate" function: the identify, means signal transmission, conversion, or transduction. A possible analog response is also shown. The input and output signals are actually not limited to two values, or to the range bounded by them, of the digital/binary **0** and **1** selected as appropriate for a specific application. The analog signals can also be considered beyond the "digital" range, if physically allowed, as shown by the broken line extensions. Chemical concentrations can only be nonnegative, but the binary **0** does not have to always be at the physical zero.

A simple model is offered by an irreversible diatomic chemical reaction described within the rate-equation approximation with the rate constant *k*, of the species *A*, of the initial concentration $A(0) = A_0$, pairwise combining to yield the product, *C*, of concentration *C*(*t*) at time $t \geq 0$, with, initially, $C(0) = 0$:

$$A + A \xrightarrow{k} C\,. \tag{1}$$

$$\frac{dA}{dt} = -2kA^2\,, \tag{2}$$

$$C(t) = \frac{A_0 - A(t)}{2} = \frac{kA_0^2 t}{1 + 2kA_0 t}\,. \tag{3}$$

We assume that the information-processing application identifies a reference value, $A_{\max}$, of the input as the digital **1**, and, in the simplest case, the physical zero as the digital **0** input. The product of the reaction constitutes the output signal used/measured at the "gate time" $t_g$. The binary values for the output are then set by the gate itself: **0** and **1** will be, respectively, $C = 0$ and

$$C_{\max} = \frac{kA_{\max}^2 t_g}{1 + 2kA_{\max} t_g}\,. \tag{4}$$

The logic-range variables, $x$ and $z$, represent the input, $A(0) = A_0$, and the output, $C(t_g)$, normalized to the "digital" range of values:

$$x = A_0 / A_{\max} , \tag{5}$$

$$z = C(t_g) / C_{\max} . \tag{6}$$

With these definitions, we have the gate-response function, shown in Figure 2,

$$z(x) = \frac{(1+2p)x^2}{1+2px} , \tag{7}$$

which depends on the parameter combination

$$p = kA_{\max} t_g . \tag{8}$$

The digital-**1** of the input, $A_{\max}$, is generally determined by the specific application or other network elements to which the present gate is connected. However, we can to some extent vary the reaction rate constant, $k$, by altering the physical and chemical conditions of the system, within the range allowed by the operational environment. We can also possibly adjust the reaction time, $t_g$. This allows a certain degree of control of the "response function shape" which can be used for gate design and optimization, as elaborated on in the following sections.

The considered chemical reaction generally yields concave shapes, shown in Figure 2. Furthermore, as illustrated in the figure, the actual shape of the gate response function cannot be varied significantly by just "tweaking" the parameters. Indeed, order of magnitude variations in the parameter values — which might not be practical in many applications — are needed to achieve qualitatively different shape. This difficulty[10,12,19] is shared by most, but especially catalytic, (bio)chemical computing gates and can be traced to that "activities" of reagents and catalysts are effectively cancelled out in the leading, linear order in defining the rescaled "logic-range" variables, see Equations (5)-(6). Finally, we note that both variables in Figure 2 need not be limited to [0,1]; the curves are well-defined and shown for $x$ and $z$ larger than 1 as well.

## 3. Noise Handling Considerations

Important topics of noise amplification and filtering will be addressed here on the example of the "identity" gate just introduced. Two-input/one-output **AND** gates will be addressed in the next section. The following reaction exemplifies a response more realistic of typical chemical kinetics:

$$A + B \xrightarrow{K} C\,, \tag{9}$$

with the rate constant $K$ and initial conditions $A(0) = A_0$, $B(0) = B_0$, $C(0) = 0$, and the output signal measured as $C(t_g)$. This reaction can be considered as a two-input process. However, here we regard $A_0$ as the input set by the environment in which the gate is used, whereas $B_0\ (< A_0)$ will be for now assumed small (so that it will limit rather than drive the output) and controllable-supply "gate machinery" chemical.

The rate equations and their solution are summarized in

$$\frac{dA}{dt} = -KAB = -KA(A - A_0 + B_0)\,, \tag{10}$$

$$C(t) = \frac{A_0 B_0 [1 - e^{-(A_0 - B_0)Kt}]}{A_0 - B_0 e^{-(A_0 - B_0)Kt}}\,. \tag{11}$$

Equations (5)-(6) are then used to rescale the input and output in terms of the "logic" variables:

$$z(x) = \frac{x(1-e^{-ax+b})(a-be^{-a+b})}{(1-e^{-a+b})(ax-be^{-ax+b})}. \tag{12}$$

This expression depends on two dimensionless combinations,

$$a = KA_{\max}t_g , \quad b = KB_0t_g . \tag{13}$$

These parameters can be controlled, see Figure 3, by changing the physical and chemical conditions (vary the rate constant *K*), the "gate machinery" chemical supply, $B_0$, and the reaction time, $t_g$.

One can prove[188] by algebraic considerations that the function in Equation (12) is always monotonically increasing *convex*; see Figure 3. Indeed, for catalytic (bio)chemical reactions and many other (bio)chemical processes, convex response curves — and surfaces, for more than one input — are generally expected. The product of the reaction — the output — is typically proportional to (linear in) the input-signal chemical concentration(s) for *small* inputs. For *large* inputs, the output is usually limited, for example, by the reactivity of the available (bio)catalyst, or, in our case, the availability of the second reactant, *B*. Therefore, the output signal reaches saturation in the large-input limit.

There are many possible sources of error in gate functioning. The most obvious noise is that in the inputs, which is actually quite large in chemical and biochemical environments. The gate function will transfer the resulting distribution of the input values into noise in the output. In addition, the binary **0** and **1** signal values need not be sharply defined. In applications, input/output signals in certain ranges of values may sometimes constitute **0** or **1**. For example, a range of normal physiological concentrations can be **0**, whereas an interval of pathophysiological values can be **1**, and these ranges need not even be bounded. The gate function can also be noisy, and the distribution of its values can be displaced away from the desired digital values/ranges: In our notation, noise and fluctuations in concentrations and physical parameters of the system can lead to a distribution of the values of $z(x)$, for each *x*, rather than a sharply defined function such as in Equations (7), (12). The mean values of this distribution need not pass precisely through the expected logic values at the "logic" inputs.

We will term "analog" the noise due to the spread of the output signal about the reference "digital" values (or ranges). In order to prevent buildup of noise as gates and other network elements are connected, we have to pass our signals through "filters" with response close to that shown in Figure 4. Ideally we would like to have the sigmoidal property — small slopes/gradients at and around the digital values — in all or most of our gates. Filters can also be used as separate network elements. There is evidence that filtering for suppressing analog noise buildup is utilized by Nature.[189,190] Experimental attempts to realize a biochemical filter have only recently shown preliminary successes.[191]

The inset in Figure 4 points out the property that filtering can push those values which are far from the correct digital result even closer to the wrong answer. Thus, the process of digitalization itself introduces also the "digital" type of noise: small probability of a wrong binary value. Digital errors are not very probable and only become important to correct for large enough networks. Standard techniques based on redundancy are available[192,193] for digital error correction.

For enzyme-based gates studied by our group, for the presently realized network sizes the analog error correction is important and has recently received significant attention.[10,12,14,15,17,19,20,30] It has been estimated[10,12] that up to order 10 such gates can be connected in a network before digital error correction is warranted.

Experimental realizations of the sigmoidal behavior (Figure 4) have been an ongoing effort,[188] based on the ideas[10,188,190] that an additional reactant, *F*, which depletes the product, but can only consume (react fast with) a small quantity of it, will suppress the response at small inputs without voiding the saturation property at large inputs, thus yielding a sigmoidal response. In connection with the system of the type defined in Equation (9), we can consider reactions $C+F\xrightarrow{\rho}\ldots$, with a fast rate, $\rho$, and with $\ldots$ denoting inert chemicals. This added reaction, however, delays the saturation at large inputs. Another option is $C+F\xrightarrow{\rho}A+\ldots$, which, however, introduces a feedback loop — the effects of which have not been studied — by regenerating some of the input. A variant realized experimentally[188] was for a system more complicated than the single reaction in Equation (9), and then the output of the added process, $C + F \xrightarrow{\rho} S + \ldots$, included an intermediate product, *S*, involved in the last reaction step that leads to the output signal. The closest equivalent for the system of Equation (9) would be to have $C + F \xrightarrow{\rho} B + \ldots$. Interestingly, the system of rate equations obtained by adding this reaction to the one in Equation (9), is still solvable in a closed form in quadratures, because the solution steps lead to a single differential equation which, while nonlinear, is of the Bernoulli form.[194] However, the expressions obtained are sufficiently complicated so that the closed-form results are unilluminating, and numerical evaluation is needed to make them illustrative/tractable, which is outside the scope of the present review.

Ultimately, the output signal of several (bio)chemical information processing steps in near-term sensor applications of the "decision-making" type,[14,15] will likely be fed into conventional electronic circuitry. Well developed research, not reviewed here, addresses the interfacing of biomolecular logic with "smart" signal-responsive[16,59,74,101,195-205] materials and with electrodes and bioelectronic devices.[16,206-212] The interfacing/transduction of (bio)chemical signals to electronic ones, can also incorporate a filtering "sigmoidal" property, as has been recently experimentally demonstrated.[16]

The above discussion reveals that while (bio)chemical filtering is a desirable property for gates and network elements, its direct experimental realization has thus far been quite limited. Therefore, efforts has been devoted to directly minimizing noise amplification for gates with *convex* response curves/surfaces of the "standard" (bio)chemical-reaction type. For single-input/single-output gates, such as the illustrated "identity" function, Figure 3, noise amplification (increase in the spread of the noise distribution due to the gate function) is simply related to the maximal of the two slopes at the binary points, and it can be minimized by having both slopes close to 1, i.e., a nearly straight line response curve (see Figure 3).

However, a danger — also identified in designing filtering systems[188] — with this approach has been that the near-linear behavior is realized straightforwardly when the reaction is far from saturation. The latter regime corresponds to weak output signal, and therefore while there is no noise *amplification*, another source of relative sensitivity to noise is introduced: that of the random "environmental" external noise being comparable to the spread of the binary **0** and **1** signal reference values.

One solution has been to "drive" the reaction by flooding the system with reagent(s) that will increase the process rates. For example, for the reaction scheme considered in this section, Equations (9)-(13), we could effectively increase the overall time-dependent rate, $KB(t)$, of the consumption of the input, $A$, by selecting $B_0 >> A_0$ (instead of the originally assumed regime $B_0 < A_0$) for all the relevant inputs: With $B_0 >> A_{\max}$, we have $b >> a$, and the response curve, Equation (12), then reduced to the linear one, $z(x) \approx x$ (for all $x \in [0,1]$). This is perhaps not an interesting "curve" to consider, but it does the job of avoiding/minimizing noise amplification. The situation with the two-input/one-output gate functions is more complicated and is discussed in the next section.

## 4. Optimization of The AND Gates

**AND** is the most studied gate in the (bio)chemical computing literature, and practically the only one explored in detail for the optimization of its response. Since the truth table for the **AND** gate is that the output **1** is obtained *only* when *both* inputs are **1**, **AND** is a natural outcome as a product of a two-input chemical reaction. The **AND** gates themselves are not universal, but they become such if supplemented by **NOT**: **NAND** (not-**AND**) gates can be networked to realize an arbitrary binary function. Indeed, the **NOT** version of filtering[16] — the vertically flipped sigmoidal curve as compared to Figure 4 — would be particularly interesting to realize and widely incorporate in networked biochemical processes.

We consider a simple model for the **AND** gate in chemical computing: We now regard the reaction in Equation (9): $A + B \to C$, as a two-input, one-output process. We introduce the "logic-range" variable for the input $B$, rescaled to the binary interval [0,1],

$$y = B_0 / B_{\max}, \tag{14}$$

paralleling the definitions in Equations (5)-(6). Here $B_{\max}$ is the reference value for logic-**1** of the *B*-input. The quantity *z* defined in Equation (6), is now a two-variable function, $z(x, y)$, describing the **AND**-gate response *surface* shape. The solution of the rate equations, given by Equation (11), is now recast in terms of the new set of the "logic-range" variables to yield

$$z(x, y) = \frac{xy(\alpha e^{\alpha} - \beta e^{\beta})(e^{\alpha x} - e^{\beta y})}{(e^{\alpha} - e^{\beta})(\alpha x e^{\alpha x} - \beta y e^{\beta y})}, \tag{15}$$

where we defined

$$\alpha = KA_{\max} t_g, \quad \beta = KB_{\max} t_g. \tag{16}$$

These are similar to the (dimensionless) parameters in Equation (13), but we now have less control over their values, because their ratio is set by the application (the environment) of the gate which in most cases predetermines both $A_{\max}$ and $B_{\max}$. Only the product $Kt_g$ can be externally adjusted.

The response surfaces are illustrated in Figures 5 and 6. Recall that the noise in the input signals is passed to the output, with the added noise effects due to the gate function: imprecise (on average) and fluctuating values of $z(x, y)$. In addition to designing gates as precise as possible, we can also minimize the propagation of analog noise, and hopefully avoid noise amplification, by finding parameters (such as $Kt_g$) that yield gates which suppress, or at least diminish spread in the input signals, by having small slopes near the logic points. Indeed, the absolute value of the gradient vector, $\left|\vec{\nabla} z(x, y)\right|$, calculated at the logic points, measures the noise spread amplification or suppression. This is only relevant if the input noise distribution is narrow and also provided the gate function $z(x, y)$ is smooth in those regions near the logic points which are approximately the size of the spread of the noise distributions. Relatively smooth $z(x,y)$ shapes are illustrated in Figure 5. We can try to identify parameter values for which the *largest* of the four gradients, $\left|\vec{\nabla} z(x,y)\right|_{x=0,y=0}$, $\left|\vec{\nabla} z(x,y)\right|_{x=1,y=0}$, $\left|\vec{\nabla} z(x,y)\right|_{x=0,y=1}$, $\left|\vec{\nabla} z(x,y)\right|_{x=1,y=1}$ is as small as possible (note that $\left|\vec{\nabla} z\right|_{00}$ is always zero for this particular model). For this calculation, let us for now assume that both $\alpha$ and $\beta$ can be adjusted independently. By numerical calculation we then find that for moderate values of $\alpha$ and $\beta$, the minimum is obtained for $\alpha = \beta \approx 0.4966$, and is given by $\left|\vec{\nabla} z\right|_{10} = \left|\vec{\nabla} z\right|_{01} = \left|\vec{\nabla} z\right|_{11} \approx 1.1796$.

It turns out that gate functions of this type amplify analog noise even under optimal conditions. The noise amplification in the best case scenario is about 18%. Studies[10,12,19] of enzyme-based **AND** gates, which have utilized more realistic (and thus more complicated and not exactly solvable) rate-equation models appropriate for biocatalytic reactions, found similar estimates. Experimental data were fitted and results were numerically analyzed by using both the rate equation approach and more phenomenological shape-fitting forms, the latter exemplified in the next section. If not optimized, smooth, convex gates corresponding to typical (bio)chemical reactions can have very large noise amplification, 300–500%. Reaching the optimal conditions is not always straightforward primarily because the gate function shape depends only weakly on parameter values. Even under optimal conditions, at least about 20% noise amplification is to be expected.

For fast enough reactions the maximal gradient can be smaller than ~1.2 and even decrease below 1, which would suggest noise suppression. However, as illustrated in Figure 6, the gate function surface then develops sharp features, and the gradients can no longer be used as measures of noise amplification, because they remain close to the logic-point values only in tiny regions near these points, as compared to the noise spread of at least several percent typical for (bio)chemical signals. Generally when the spread of the noise is larger than the $x$ and $y$ scales over which the gate function or its derivatives vary significantly, one can assume a certain shape of the *input* noise distribution, such as a product of approximately Gaussian distributions in $x$ and $y$ for inputs at each of the logic points, or half-Gaussian, if the logic zero is exactly at the physical zero. Given a model for the gate response function, one can then numerically calculate[10] the *output* signal distribution for each of the inputs and thus estimate the noise amplification factor.[10,12,14]

The "ridged" gate response function (e.g., Figure 6) was first encountered[12] in a study of an enzymatic system which also realized a smooth-response counterpart when a different chemical was used as one of the inputs.[12] The reaction kinetics was more complicated than in the present model, but the finding has confirmed the general expectations: The optimal conditions are obtained with a symmetrically (diagonally) positioned ridge, as in Figure 6, and the noise amplification factor is then only slightly larger than 1, estimated by considering distributions. Thus, noise amplification is practically avoided. However, such gates do not have the noise-suppression (filter) property.

Figure 7 presents a schematic of an **AND** gate response sigmoidal in only one of the two inputs, which was recently explored and experimentally realized.[14] Many allosteric enzymes have such a "self-promoter" property with respect to one of their substrates (input chemical species). A key finding[14] has been that the single-sided sigmoidal shape can be tuned by parameter adjustment to have the noise amplification factor only slightly above 1, so that there is practically no noise amplification. However, a desirable two-sided sigmoid response, also shown schematically in Figure 7, has not to our knowledge been realized at the level of a single **AND** gate, in chemical or biomolecular computing literature. Certain biochemical processes in nature, which are much more complex than our synthetic **AND**-gate systems, do realize[189] a two-sided sigmoidal response.

## 5. Networking of AND Gates

We have seen that optimization of (bio)chemical gates one at a time is not straightforward. In most cases a rather large variation of the controllable parameters is needed: physical and chemical conditions, reactant concentrations and in some cases choice of (bio)chemical species, which may not be experimentally feasible. The actual detailed kinetic modeling of the reactions involved, especially for biomolecular systems, is in itself a challenging task:[10,12,14,15,17] The kinetics of most biomolecular processes, specifically those used for **AND** gates, is complex and not well studied. The quality of the experimental data for the gate-response function is limited due to the noise in the gate-function itself, limited life-time for constant activity of the biocatalytic species, etc. Thus, multi-parameter complex reaction schemes are difficult to substantiate by data fitting in the gate-design context which requires models to work for a large range of adjustable parameters.

An alternative approach involves optimization of the *relative* gate functioning in a network, whereby each gate is modeled within a very approximate, phenomenological curve/surface-fitting approach. These ideas have recently been tested[19] for coupled enzymatic reactions which include steps common in sensor development[213] for maltose and its sources. A modular network representation of the biocatalytic processes involved, is possible in terms of three **AND** gates; see Figure 8. This "cartoon" representation is actually approximate, because it obscures some of the complexity of the constituent processes.[19]

The approach involves first proposing a phenomenological fitting function for the gate response surface in terms of as few parameters as possible, enough to capture the expected *qualitative* features of the shape. For a typical convex "identity" gate, the fitting function is conveniently written as

$$z(x) \approx \frac{sx}{(s-1)x+1}. \tag{17}$$

This is a single-parameter, $s$, rational form that "looks" qualitatively correct, provided we assume

$$s > 1. \tag{18}$$

Indeed, the curve is then convex and has slope $s$ at $x, z(x) = 0,0$, and $1/s$ at $x, z(x) = 1,1$.

For each **AND** gate, we then use the two-parameter, $s > 1$ and $u > 1$, product

$$z(x, y) \approx \frac{(sx)(uy)}{[(s-1)x+1][(u-1)y+1]} . \tag{19}$$

The gradient values are $\left|\vec{\nabla} z\right|_{00} = 0$, $\left|\vec{\nabla} z\right|_{10} = u$, $\left|\vec{\nabla} z\right|_{01} = s$, and $\left|\vec{\nabla} z\right|_{11} = \sqrt{s^{-2} + u^{-2}}$. The minimum of the largest of the last three values is obtained for $s = u = \sqrt[4]{2} \approx 1.189$, which is also the *value* of the gradient, consistent with the earlier reported empirical expectation that smooth convex **AND** gates can typically be optimized at best to yield noise amplification somewhat under 20%.

Having introduced our approximate fitting functions, we now experimentally vary selective inputs in the network; see Figure 8. In the experiment,[19] each of the three inputs $x_{1,2,3}$ was separately varied between 0 (corresponding to the binary **0**) and the reference value pre-defined as **1**, while all the other inputs (including $y_3$) where at their reference **1** values. In fact, when the parameterization of Equation (19) is applied to all three gates in our network of Figure 8, we get a rational expression for $z$ as a function of all the four inputs ($x_{1,2,3}$ and $y_3$). Setting all of them but a single $x$-input to 1, we get the parameterization for the measurement with that input varied; we only keep that varying argument of $z(\cdots)$ for simplicity:

$$z(x_1) = \frac{s_1 x_1}{(s_1 - 1)x_1 + 1} , \tag{20}$$

$$z(x_2) = \frac{s_2 u_1 x_2}{(s_2 u_1 - 1)x_2 + 1} , \tag{21}$$

$$z(x_3) = \frac{s_3 u_1 u_2 x_3}{(s_3 u_1 u_2 - 1)x_3 + 1} . \tag{22}$$

Interestingly, each data set only depends on a single parameter ($s_1$, $s_2 u_1$, or $s_3 u_1 u_2$).

While we only get partial information on the gate functioning, we can attempt to "tweak" the relative gate activities in the network to improve the stability. If the proposed approximate description is semi-quantitatively accurate for a given gate, then the parameters $s$ and $u$ for that gate will be functions of adjustable quantities, such as the gate time, input concentrations of some of the chemicals, and reaction rates (which can in turn be controlled by the physical and chemical conditions). In addition, $s$ and $u$ can depend on other quantities which are not controllable.

Without detailed rate-equation kinetic modeling this parameter dependence is not known. However, examination of the fitted quantities ($s_1$, $s_2u_1$, $s_3u_1u_2$) still provides useful information on the *relative* effect that the gates have, by their contribution to the gradients at various logic points, when compared to the optimal values ($s_1 = 2^{1/4}$, $s_2u_1 = 2^{1/2}$, $s_3u_1u_2 == 2^{3/4}$). The initial sets of data[19] were collected with the experimentally convenient, randomly selected values for the "gate machinery" and other parameters. Examination of the results[19] has lead to a semi-quantitative conclusion that the deviations form the optimal values could largely by attributed to the gate which is the closest to the output in Figure 8 ($z = x_1$ **AND** $y_1$): it was too "active" as compared to the other two gates (means, its biocatalytic reaction was too fast). A new experiment was then devised[19] with the concentration of the enzyme catalyzing this gate's function reduced by an order of magnitude (approximately 11 fold). The new data collected for the modified network, when fitted, yielded $s_1$, $s_2u_1$, $s_3u_1u_2$ values significantly closer to the optimal.[19]

## 6. Conclusions and Challenges

We reviewed aspects of and approaches to gate optimization for control of the analog noise amplification, which is important for connecting gates in small networks. For larger networks, digital error correction by redundancy will also have to be implemented, and various network elements will have to be devised for filtering, signal splitting, signal balancing, gate-to-gate connectivity, memory, interfacing with external input, output and control mechanisms, etc.

We used simple rate-equation models which allow exact solvability, to illustrate and motivate the discussion. Thus, we avoided presenting experimental data and their numerical analysis, which can be found in the cited articles, while various chemical and biochemical gate examples are offered in other reviews in this Special Issue. Our presentation has been limited to **AND** gates and related systems. Indeed, all the recent studies, with one exception: an **XOR** gate,[214] of noise control in (bio)chemical computing have thus far been for **AND** gates and, furthermore, again with just one recent exception,[215] only for those with the binary **0** set at the physical zeros of chemical concentrations. While these limitations are natural for chemical kinetics, they are definitely not typical for applications envisaged, notably, multi-input biomedical sensing.[15]

As new experiments on mapping out (bio)chemical gate functioning and network designs are reported, new features of noise and error control will be explored. Specifically, noise in the gate function itself, including spread of its values and imprecise mean-values — not exactly at the expected reference output **0** or **1**, with deviations possibly also different for various inputs that should ideally yield the same logic output — will have to be considered and corrected, most likely by filtering. Indeed, we conclude by emphasizing that, while longer-term, network design and scaling up will be crucial, the shorter-term challenges in (bio)chemical information processing have been to design and experimentally realize versatile and effective *(bio)chemical filter processes* and other non-binary network elements that can be concatenated with various binary logic gates.

**Acknowledgements**

The author acknowledges research funding by the United States National Science Foundation (grant CCF-1015983).

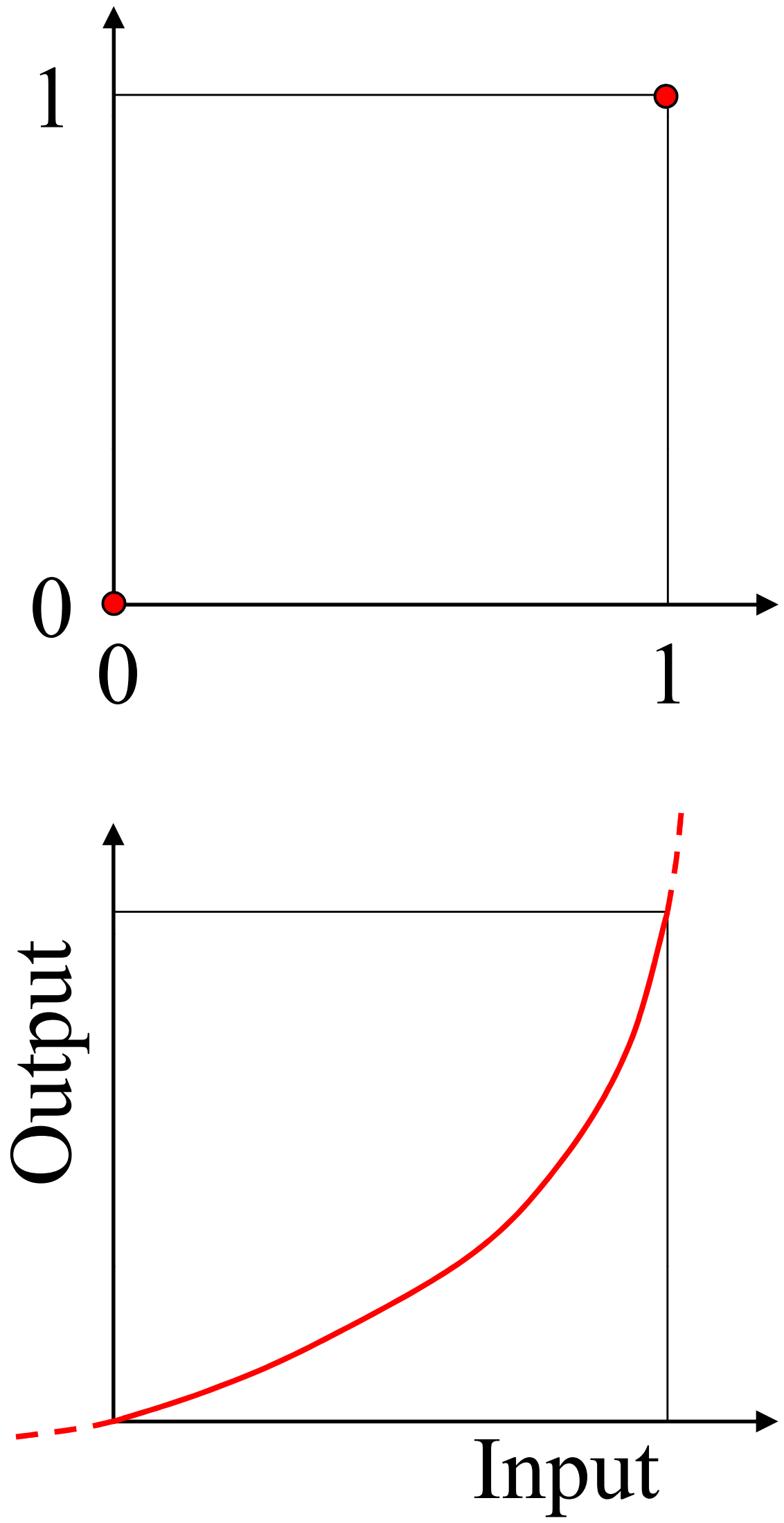

**Figure 1.** *Top:* The identity binary function: digital **0** and **1** are mapped to the same values. *Bottom:* A possible response curve.

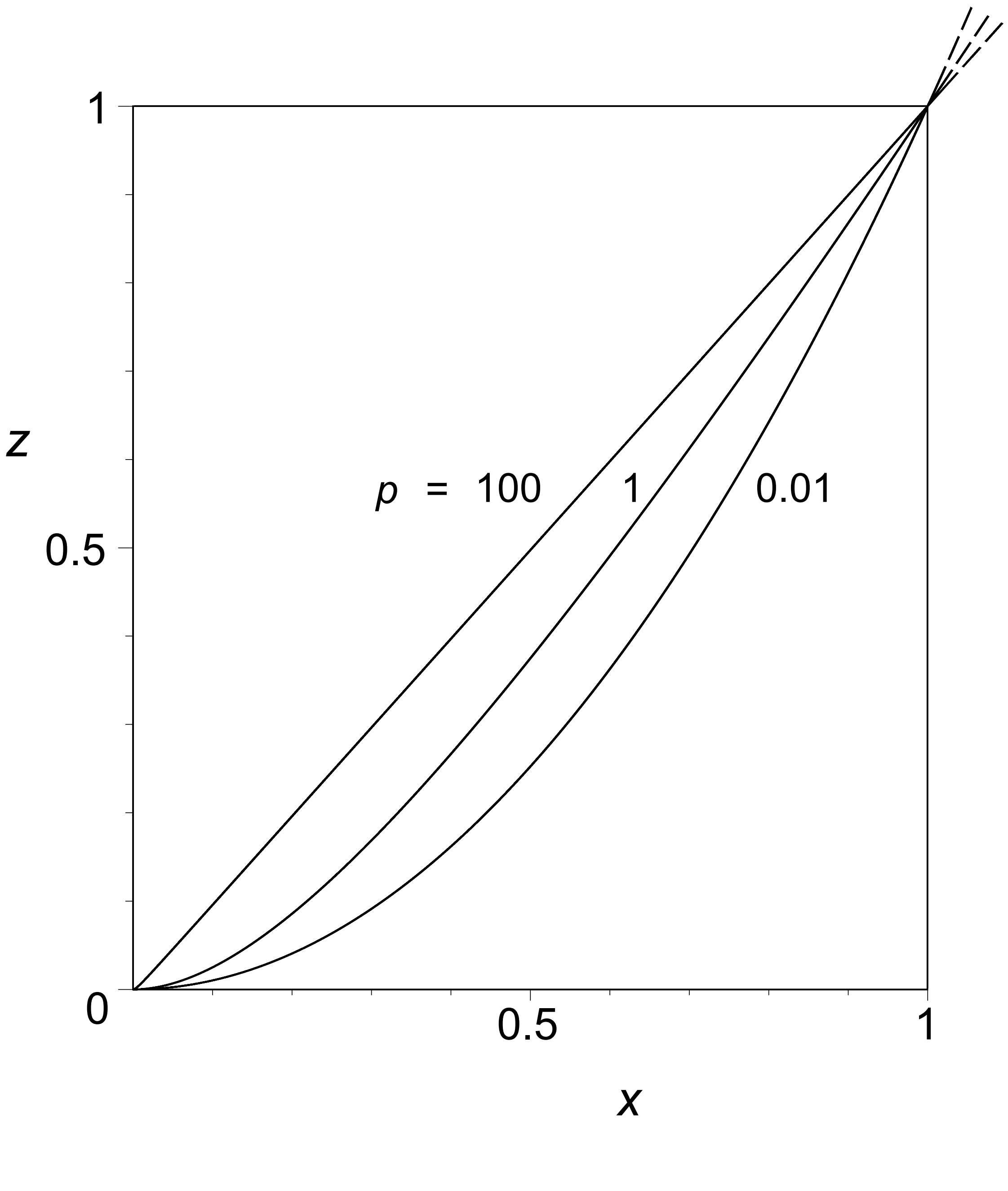

**Figure 2.** The response function for the reaction $A + A \to C$, see Equations (7)-(8), for three values of the parameter *p*. All the curves are concave.

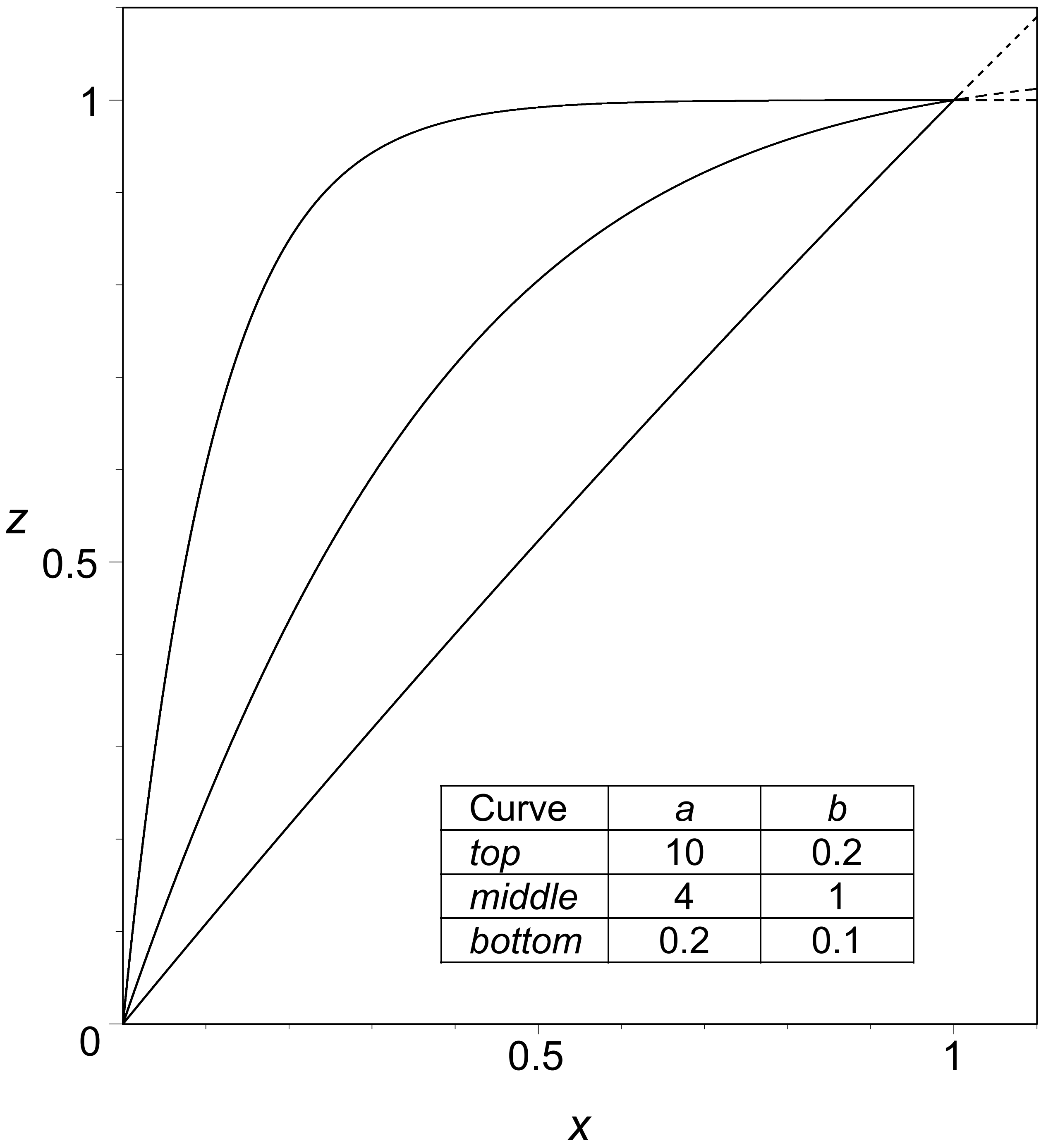

**Figure 3.** The convex response function, Equation (12), for the reaction $A + \cdots \rightarrow C$, where the omitted reactant is not considered a variable input, here shown for three different choices of the parameters *a* and *b*.

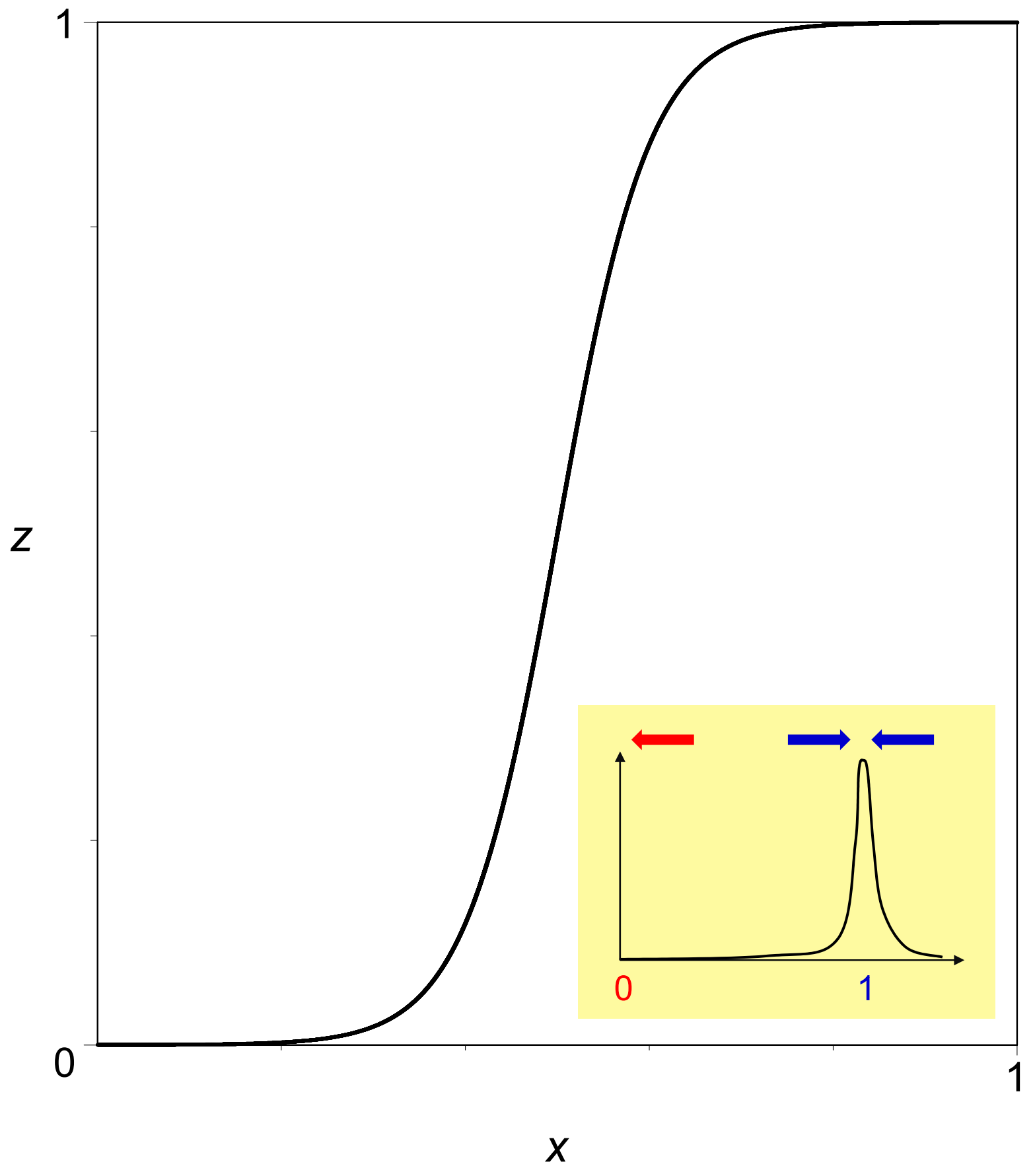

**Figure 4.** Sigmoidal response for the "identity" gate to act as a filter. The central inflection region should be narrow and positioned away from both logic **0** and **1**, and the slopes near both binary values should be zero or very small. Here we assume that the signals can only be nonnegative; the sigmoidal curve continues smoothly (not shown) past the point $x, z = 1, 1$. The *inset* illustrates a possible input, $x$, distribution, here spread about **1** — assumed to be the input/output digital value. The filtering will drive the output, $z$, values originating from $x$ close to the peak of the distribution, towards the correct digital answer, **1** (shown by the pair of facing arrows). The values in the tail of the distribution will be driven towards the wrong digital answer, **0** (shown by the unpaired arrow); this results in a small-probability "digital" error. Similarly, distributions peaked near **0**, when it is the expected digital value (not shown), will also be sharpened, but the tail values will be driven to the wrong digital answer.

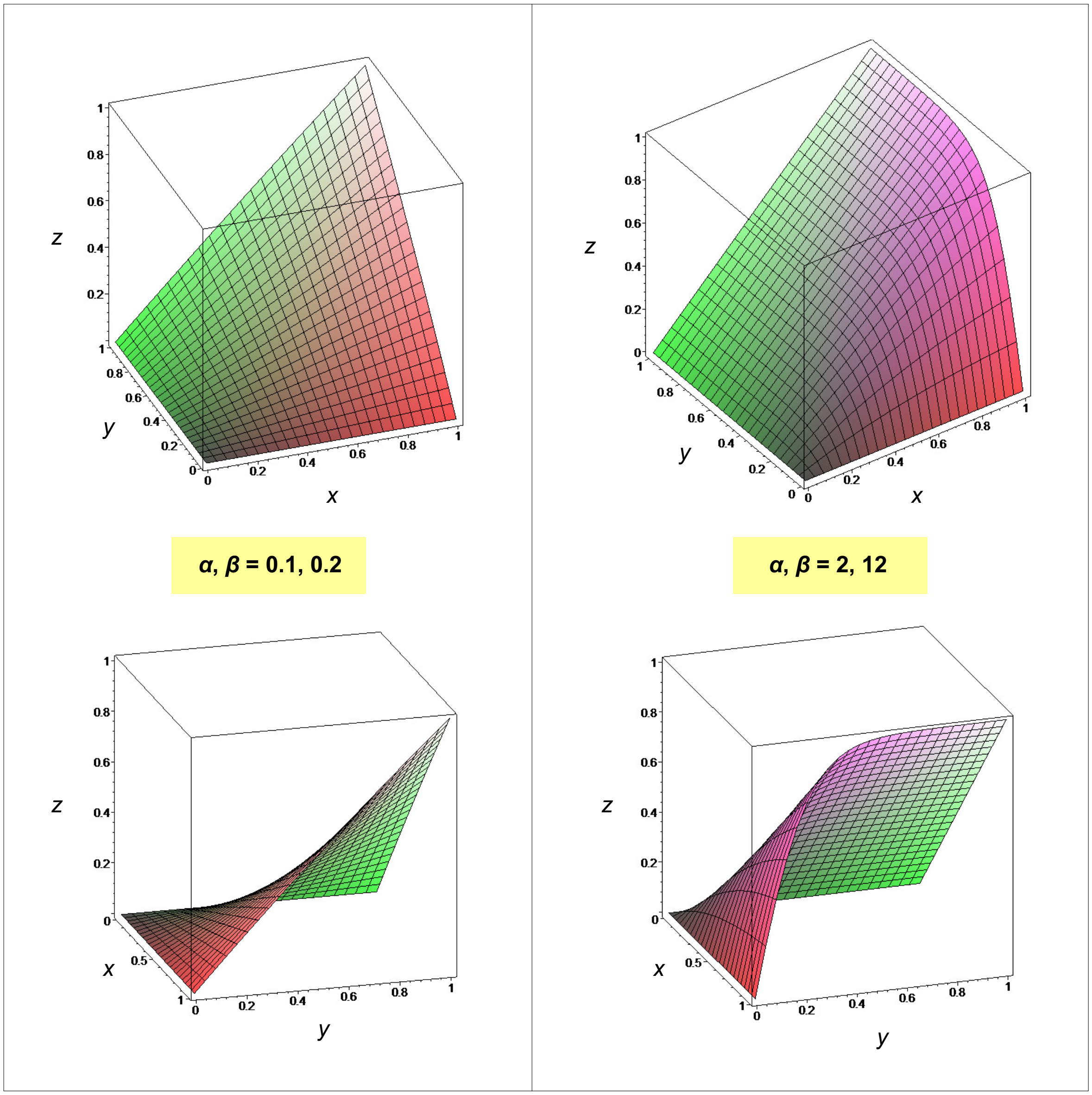

**Figure 5.** Examples of smoothly varying response surfaces, $z(x,y)$, for the **AND**-gate function representing the reaction $A+B \to C$, Equations (15) and (16), for two choices of the parameters $\alpha$ , $\beta$ . The *top* images give the frontal view, whereas the *bottom* images offer the back view for the two surfaces.

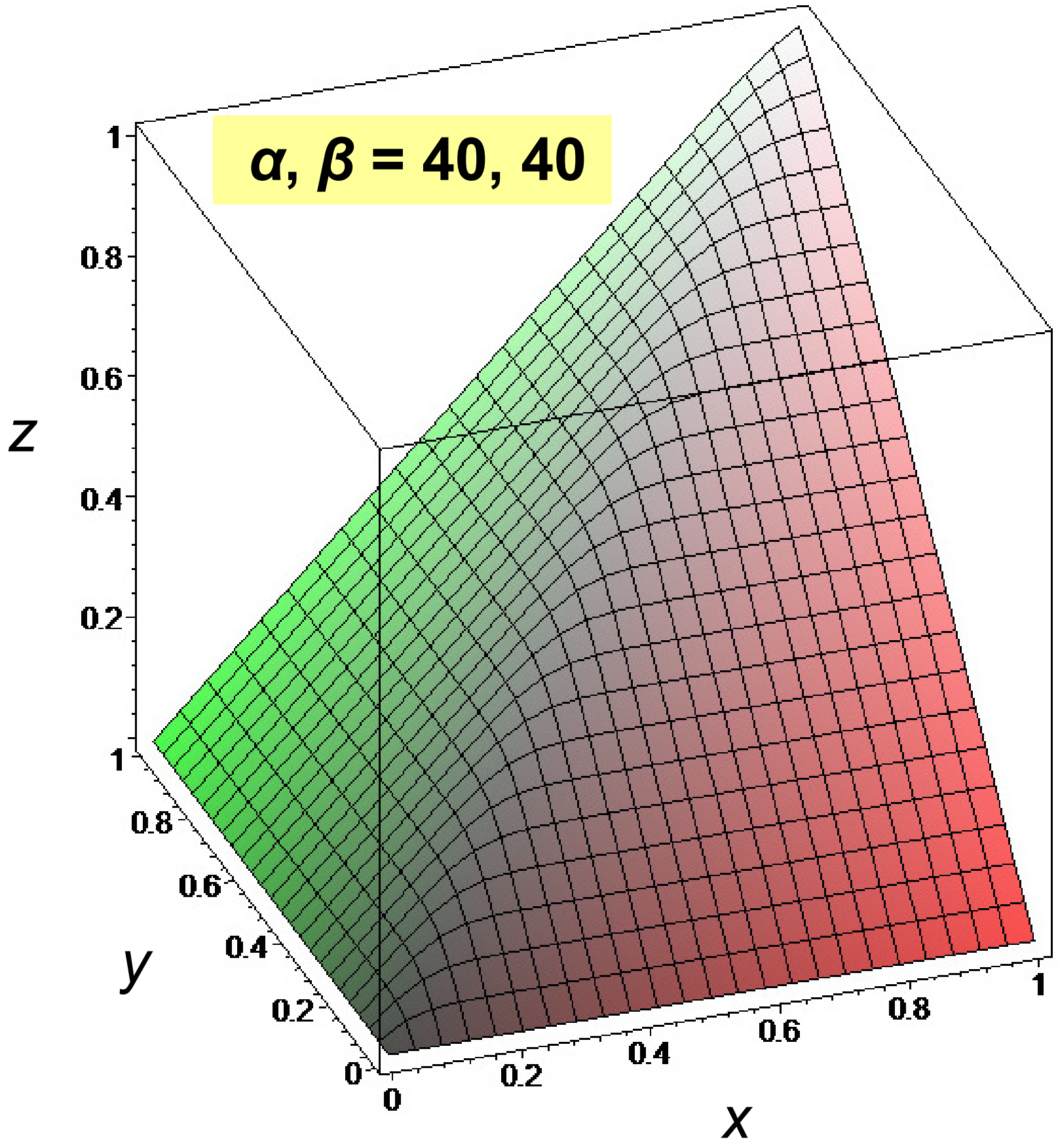

**Figure 6.** Same as in Figure 5, but for the fast-reaction case (large, here equal, $\alpha, \beta$). The emergence of a response surface with non-smooth features is seen: formation of a ridge (here, along the diagonal), and also the shrinking of the region for which the value of the gradient near the point (0,0) remains small. Note that the gradient *at* the origin, $\left|\vec{\nabla} z(x,y)\right|_{x=0,y=0}$, is zero for all three surfaces shown in both figures. Nonuniformities also set in along the ridge region, including near the point (1,1). A forming, off-diagonal ridge can already be seen in the right images in Figure 5, due to a relatively large value of $\beta$ there.

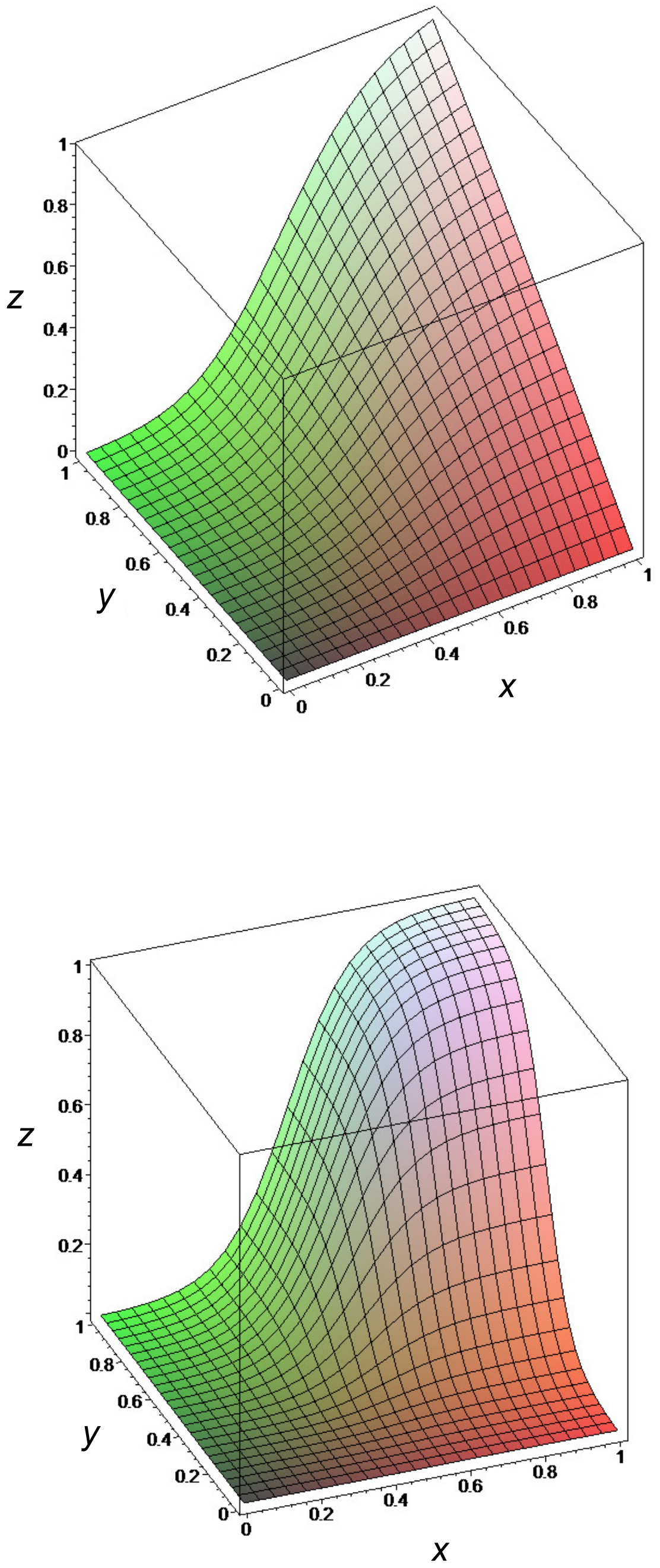

**Figure 7.** *Top:* Schematic of a one-sided sigmoidal behavior. *Bottom:* A desirable two-sided sigmoidal response for **AND** gates.

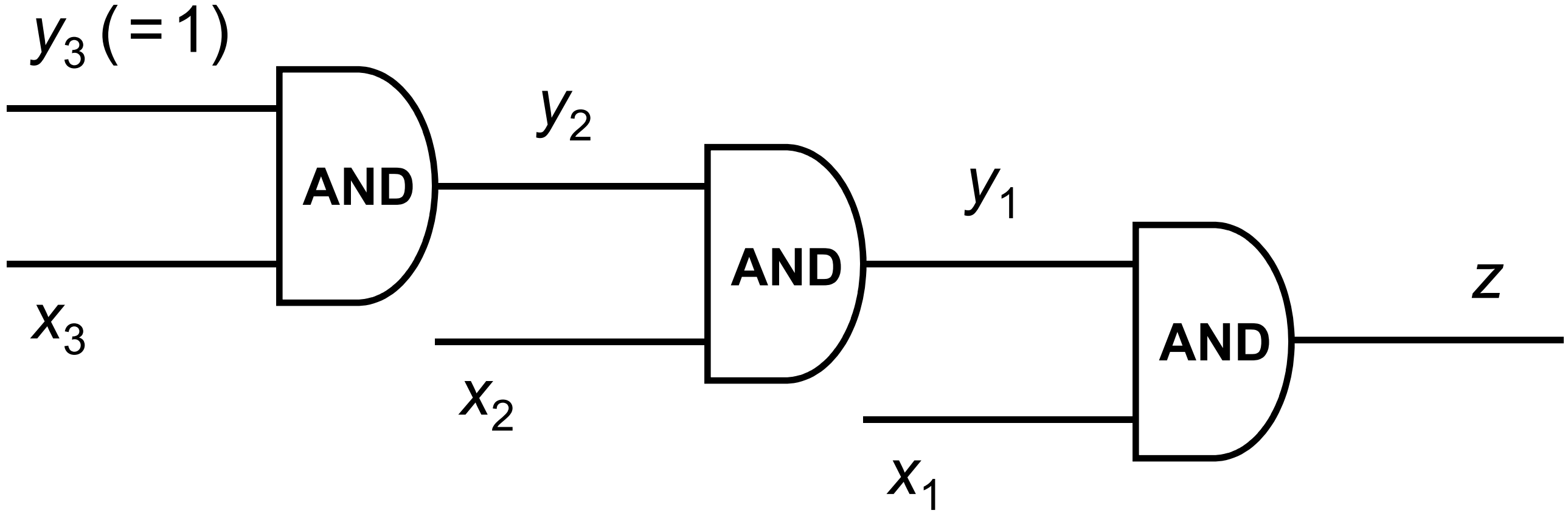

**Figure 8.** The three-gate network,[19] with varied inputs $x_{1,2,3}$, and constant $y_1$.